\documentclass[11pt,a4paper]{article}
\pdfoutput=1
\usepackage{jheppub}

\usepackage{multirow,bigdelim}

\usepackage[utf8x]{inputenc}
\usepackage[T1]{fontenc}
\usepackage{textcomp}

\usepackage{lmodern}

\usepackage{lineno}

\usepackage{float} 
\usepackage{subfig} 
\usepackage{amsmath} %
\usepackage{amstext}

\usepackage{graphicx}%
\usepackage{tabularx,ragged2e}%

\newcolumntype{Y}{>{\centering\arraybackslash}X}

\newcommand{\eqs}[1]{\begin{equation} \begin{split} #1\end{split} \end{equation} }

\def\ie{{\it i.e.}}
\def\eg{{\it e.g.}}

\def\GeV{{\rm GeV}}
\def\GeV2{{\rm GeV}^2}

\renewcommand{\P}{{\cal P}}

\newcommand{\ce}[1]{Eq.~(\ref{#1})}
\newcommand{\cf}[1]{{Fig.~\ref{#1}}}
\newcommand{\ct}[1]{{Table~\ref{#1}}}

\title{Associated production of a quarkonium and a $Z$ boson at one loop in a 
quark-hadron-duality approach}
\author[a]{Jean-Philippe Lansberg}
\author[b]{Hua-Sheng Shao}
\affiliation[a]{IPNO, CNRS-IN2P3, Univ. Paris-Sud, Universit\'e Paris-Saclay, 
91406 Orsay Cedex, France}
\affiliation[b]{Theoretical Physics Department, CERN, CH-1211, Geneva 23, 
Switzerland}

\emailAdd{Jean-Philippe.Lansberg@in2p3.fr}
\emailAdd{huasheng.shao@cern.ch}

\abstract{
In view of the large discrepancy about the associated production of a prompt 
$J/\psi$ and a $Z$ boson between the ATLAS data at $\sqrt{s}=8$ TeV 
and theoretical predictions for Single Parton Scattering (SPS) contributions, 
we perform an evaluation of the corresponding cross section at one loop accuracy
 (Next-to-Leading Order, NLO) in a quark-hadron-duality approach, also known as 
the Colour-Evaporation Model (CEM). This work is motivated by (i) the extremely 
disparate predictions based on the existing NRQCD fits conjugated with the 
absence of a full NLO NRQCD computation and (ii) the fact that we believe
that such an evaluation provides a likely upper limit of the SPS cross section. 
In addition to these theory improvements, we argue that the ATLAS estimation of 
the Double Parton Scattering (DPS) yield may be underestimated by a factor as 
large as 3 which then reduces the size of the SPS yield extracted from the ATLAS 
data. Our NLO SPS evaluation also allows us to set an upper limit on 
$\sigma_{\rm eff}$ driving the size of the DPS yield.
Overall, the discrepancy between theory and experiment may be smaller
than expected, which calls for further analyses by ATLAS and CMS, for which 
we provide predictions, and for full NLO computations in other models.
As an interesting side product of our analysis, we have performed the first NLO 
computation of $d\sigma / dP_T$ for prompt single-$J/\psi$ production in the CEM
 from which we have fit the CEM non-pertubative parameter at NLO using the most 
recent ATLAS data.
}
\date{\today}

\begin{document}

\maketitle

\section{Introduction.}

With the advent of the LHC, the observation of the associated production of a quarkonium 
and a vector boson became possible. Pioneering studies of CDF at 
the Fermilab-Tevatron~\cite{Acosta:2003mu,Aaltonen:2014rda} were motivated by 
the search for a charged $H^\pm$ decaying in a pair of $\Upsilon+W^\pm$ for 
instance. In 2014, ATLAS succesfully observed for the first time the simultaneous 
production of $J/\psi+W$~\cite{Aad:2014rua} and $J/\psi+Z$~\cite{Aad:2014kba}. 
As for now, the similar processes involving bottomonia have never been 
observed\footnote{Another class of reactions which is of interest is that of 
the production of a quarkonium + a photon. It is has been proposed to constrain the 
quarkonium-production mechanisms~\cite{Roy:1994vb,Mathews:1999ye,Li:2008ym,Lansberg:2009db,Li:2014ava},  
to measure different characteristics of the gluon content of the 
proton~\cite{Doncheski:1993dj,Dunnen:2014eta} as well as to probe the $H^0$ 
coupling to the heavy-quarks~\cite{Doroshenko:1987nj,Bodwin:2013gca}. This was the motivation of
the sole experimental study of this channel by ATLAS~\cite{Aad:2015sda} but 
it mainly focused  on the search for a signal from a $H^0$ decay rather than to the 
QCD continuum (no cross-section extraction was performed) which interests us here.} 

As its customary with quarkonium physics, these new measurements brought to 
light significant --to say the least-- tensions with the existing theories.
Whereas it is very likely that such simultaneous productions may come from two 
independent parton scatterings -- also called Double Parton Scattering (DPS) -- 
the ATLAS collaboration concluded that the usual production from Single Parton 
Scattering (SPS) was also relevant and as a matter of fact was, according to 
their analysis, significantly above existing theoretical predictions 
(see~\cite{Mao:2011kf,Gong:2012ah} for $J/\psi+Z$ and \cite{Li:2010hc,Lansberg:2013wva} 
for $J/\psi+W$). In the case which interests us here, namely $J/\psi+Z$, the SPS 
yield extracted by ATLAS~\cite{Aad:2014kba} is more than five times larger 
(corresponding to 2-$\sigma$ deviation) than the largest of the theory evaluations 
from NRQCD (with Colour-Singlet (CS) and/or Colour-Octet (CO) 
contributions)\footnote{We noticed that the theory numbers quoted in 
\cite{Aad:2014kba} have probably been misconverted into the ratio $R$  which is 
compared to the data (see below). We have made thus sure that our computation 
of $R$ is indeed compatible with what is implied in the ATLAS 
definition~\cite{private-Stefanos}. As a check, we have recomputed the cross 
sections and $R$ in NRQCD at Leading Order (LO) and included all the possible 
feed-downs. Despite the wrong quoted numbers by ATLAS, what we found however 
ends up to be similar than the quoted ones and this does not therefore change 
the conclusion that SPS contributions are small and that the most optimistic 
NRQCD upper values are not larger than one sixth of the DPS-subtracted ATLAS 
measurements.}.

Other quarkonium-associated-production channels have also been investigated. 30 
years after the pionneering analyses of NA3~\cite{Badier:1982ae,Badier:1985ri}, 
$J/\psi$-pair production has been analysed by the LHCb~\cite{Aaij:2011yc}, 
CMS~\cite{Khachatryan:2014iia} and ATLAS~\cite{ATLAS-CONF-2016-047} 
collaborations at the LHC as well as by the D0 collaboration~\cite{Abazov:2014qba} 
at the Tevatron. They are all compatible with CS contributions only 
(known up to NLO accuracy~\cite{Lansberg:2013qka,Sun:2014gca,Likhoded:2016zmk})
at small rapidity separations, $\Delta y$, whereas they point at a significant 
DPS contributions for increasing $\Delta y$, compatible with a $\sigma_{\rm eff}$ 
below 10 mb. We guide the reader to \cite{Lansberg:2014swa} for a detailed 
discussion of these different results and to~\cite{He:2015qya,Baranov:2015cle} 
for recent LO NRQCD analyses. $\Upsilon+J/\psi$ production has also been 
observed by the D0 collaboration~\cite{Abazov:2015fbl} with a claim that the 
yield is highly dominated by DPS contributions (see \cite{Shao:2016wor} for a 
complete and up-to-date discussion of the theoretical aspects of such a reaction). 
Similar conclusions have been drawn by LHCb for $J/\psi+$charm~\cite{Aaij:2012dz} 
and $\Upsilon+$charm~\cite{Aaij:2015wpa} production although compatible with 
$\sigma_{\rm eff}$ larger than 10 mb. It is worth emphasising that the D0 and 
ATLAS $J/\psi$-pair analyses~\cite{Abazov:2014qba,ATLAS-CONF-2016-047} are the 
only for quarkonia where DPS and SPS were separated based 
on kinematical variables.

Motivated by the discrepancy uncovered by ATLAS, we perform here the very first 
complete evaluation of the SPS yield at NLO for the production of a $J/\psi$ 
associated with a $Z$ boson in $pp$ collisions under the assumption of 
quark-hadron duality which, in the case of quarkonium production, is referred to
as the Colour-Evaporation Model (CEM). Such a computation has in fact never been 
performed at LO. This allows us to question the size of the DPS yield assumed by 
ATLAS based on their $W+$ 2-jet analysis~\cite{Aad:2013bjm} and to propose a 
solution for the present puzzle with a DPS yield roughly three times larger (yet 
compatible with the 1-$\sigma$ upper value set by ATLAS).
 
As a side product of this complete analysis, we provide another novel NLO 
analysis bearing on single-$J/\psi$ production, namely that of the 
single-$J/\psi$ $P_T$ spectrum at NLO via the computation at NLO ($\alpha_s^4$) 
of the $P_T$ spectrum of a $J/\psi$ recoiling against a parton, which we use to 
constrain the sole non-pertubative CEM parameter in a consistent way using the 
same $P_T$ region for single-$J/\psi$ data as the one for $J/\psi+Z$.  
Usually, such a parameter is fit from the total $J/\psi$ yield which is known 
at NLO ($\alpha_s^3$).

The structure of the paper is as follows. In section \ref{sec:motivation}, we 
explain the purpose of using a simple approach as the CEM in the context of 
heavy-quarkonium production. In particular, we highlight its limitation as well as 
its interest for the present study. In section \ref{sec:model}, we discuss its 
implementation in the NLO framework which we have used; we detail the computation 
of the single-$J/\psi$ $P_T$ spectrum at NLO and compare the CEM non-perturbative 
parameter which we obtained here with previous works. In section \ref{sec:result}, 
we present our results for prompt $J/\psi+Z$. We start with a discussion of the 
$P_T$-integrated yield, then discuss the $P_T$ dependence of both SPS and DPS 
contributions which finally allows us to explain why we believe that the ATLAS 
data in fact allows for a DPS yield three times as large as they initially 
assumed. Section \ref{sec:conclusion} gathers our conclusions.

\section{Why the Colour-Evaporation Model ?\label{sec:motivation}}

In parallel to the aforementioned advances in the study of quarkonium 
{\it associated} production, more precise data for {\it single}-quarkonium production
are flowing from the LHC, including cross-section measurements at larger $P_T$, 
improved measurements of feed-down fractions, more reliable multi-dimensional 
polarisation measurements (see~\cite{Andronic:2015wma} for a recent review) and, 
last but not least, the first measurement of the production cross-section of 
the spin-singlet $\eta_c$~\cite{Aaij:2014bga}.  At the great surprise of some, 
such $\eta_c$ data happened to be very well described by 
postdictions~\cite{Butenschoen:2014dra,Zhang:2014ybe,Han:2014jya} with the sole 
CS contributions (also called CSM) leaving nearly no room for CO contributions 
in this channel. Such a constraint, translated to the $J/\psi$ case via 
Heavy-Quark-Spin Symmetry (HQSS), could only be met by assuming a rather small 
value of the LDMEs for $ \langle {\cal O}_{J/\psi}({^1S^{[8]}_0)}\rangle$, which in turn induces, 
in order to reproduce the large-$P_T$ $J/\psi$ spectra at the LHC and the Tevatron, 
a somewhat large value for  $ \langle {\cal O}_{J/\psi}({^3S^{[8]}_1)}\rangle$, still in the 
ballpark of the NRQCD Velocity-Scaling Rules (VSR)~\cite{Zhang:2014ybe}. However, 
these constraints are completely at odds with the assumption made in some recent 
works \cite{Faccioli:2014cqa,Bodwin:2014gia} of a very large 
$\langle {\cal O}_{J/\psi}({^1S^{[8]}_0)}\rangle$ in order to obtain an unpolarised $J/\psi$ 
yield at large $P_T$. It is also going against the trend obtained with the 
global-fit approach of \cite{Butenschoen:2011yh} including hadroproduction data 
at lower $P_T$ as well as $\gamma p$, $\gamma \gamma$ and $e^+e^-$ data (which 
however fails to describe the polarisation of large-$P_T$ $J/\psi$). As noted 
in~\cite{Butenschoen:2014dra},  the $\eta_c$ data points at a 
$\langle {\cal O}_{J/\psi}({^1S^{[8]}_0)}\rangle$ 10 times smaller than their initial fit. 
These observations ``led [the authors of \cite{Butenschoen:2014dra} ] to conclude 
that either the universality of the LDMEs is in question or that another
important ingredient to current NLO NRQCD analyses has so far been overlooked.'' 
For instance, NRQCD, for a reason thatwe do not have uncovered yet, may not be 
reliable for polarisation observables. 

Along the same lines, it was shown in \cite{Feng:2015cba} that the NRQCD 
universality was severely challenged by the $P_T$-integrated cross sections. 
In the same work, the CEM was shown to describe reasonnably well the world data 
(see also next section). In view of this and the discussion above, it appears to 
us as legitimate to consider the CEM\footnote{In fact, as discussed in 
\cite{Bodwin:2005hm}, the CEM can be seen as a particuler -yet rather
special-- realisation of NRQCD with LDMEs following specific relations which 
however do not follow the NRQCD VSR.}  as one such models on the market allowing 
one to explore the possible production mechanisms in a rather unbiased way. The 
main advantage of using the CEM is that it is possible to perform a complete 
one-loop computation whereas NLO NRQCD computation involving all the 
relevant channnels are much more complex (without showing necessarily more stability).

Beside such encouraging comparisons of the CEM $P_T$-integrated yields, most of 
the Tevatron, RHIC and LHC data for $P_T$-differential cross sections have been 
confronted to the CEM predictions (see~\cite{Andronic:2015wma} for a representative 
selection). If one disregards the spectrum at low $P_T$  where a phenomenological 
parton-$k_T$-smearing procedure is usually applied to reproduce the cross section, 
the CEM usually tends to overshoot the data for increasing $P_T$. In order to 
address this issue, different mechanisms~\cite{Edin:1997zb,Damet:2001gu,BrennerMariotto:2001sv} 
have been considered (see~\cite{Lansberg:2006dh} for a brief overview) but none 
was the object of a consensus. In fact, the origin of the discrepancy is obvious 
and arises from the appearance at $\alpha_s^3$ of leading-$P_T$ topologies scaling 
like $P_T^{-4}$, just as those associated with the $^3S_1^{[8]}$ octet states in 
NRQCD. Whereas these were originally thought to solve the $\psi(2S)$ surplus 
at the Tevatron, recent NRQCD fits to the LHC and Tevatron data indicate that 
they have to be somewhat damped down. In NRQCD, it happens through a partial 
cancellation between both channels which show leading-$P_T$ contributions
 at NLO, namely the $^3S_1^{[8]}$ and $^3P_J^{[8]}$ states, opening the possibility 
for a dominant $^1S_0^{[8]}$ contribution in agreement with a softer $P_T$ 
spectrum\footnote{The $^1S_0^{[8]}$ channel at NLO in principle also
exhibits fragmentation-like topologies which however do not result in a $P_T^{-4}$ 
scaling since the radiated gluon from the heavy-quark pair is not soft.}.  This is 
precisely why the $\eta_c$ data are troublesome since they tend to constrain the 
importance of the $^1S_0^{[8]}$ states. In the CEM, owing to the simplicity of 
the model, no such cancellation can happen. The fragmentation contributions are 
obviously there, at any non-trivial order where the computation is carried out.

Given these discrepancies, one may wonder whether using the CEM makes sense. 
Our answer is ``yes'' as long as complete NLO predictions are not available in 
NRQCD and, most importantly, as long as the different NLO NRQCD fits on the market 
give extremely large uncertainties for quarkonium-associated-production channels. 
Even in some cases, some fits yield to negative cross sections~\cite{Li:2014ava}, 
which illustrates our current lack of understanding of the quarkonium hadronisation. 
Having a simple model at hand can then be a very useful tool to investigate tensions 
between data and experiments case by case\footnote{However, if it happens in the 
future, after confronting the CEM to a number of new observables, that the tensions 
between the $P_T$-differential cross section of single-quarkonium and the CEM are 
the only serious ones, it could mean that the underlying assumption of the CEM 
may not be too far from reality. In such a case, it could be worth devoting some 
efforts to investigate whether there are justifications and means to solve these 
tensions about the $P_T$  dependence, like an alteration of the $P_T$ spectrum by 
the feed-downs or  a kinematical bias between the computed $P_T$ of the pair and 
the one of the quarkonium eventually produced. Yet, as for now, we will consider 
the CEM as a simple and robust model which probably tends to produce a little too 
many quarkonia at increasing $P_T$ when fragmentation channels are open.}.

In view of all these arguments, we believe the CEM to be in fact currently the 
best model to investigate the ATLAS excess for $J/\psi +Z$ production since a 
complete NLO NRQCD computation is lacking and since the CEM would probably 
provide an upper theory limit given the predominance of the fragmentation 
channels for this process.

\section{Model implementation, purposes and limitations\label{sec:model}}

\subsection{The CEM and the $P_T$-integrated yields}
The CEM can be seen as the application~\cite{Fritzsch:1977ay,Halzen:1977rs} of 
the quark-hadron-duality principle to quarkonium production. The cross sections 
for quarkonium production are obtained by integrating the cross section to produce 
a $Q \bar Q$ pair in an invariant-mass region where its hadronisation into a 
quarkonium is likely. In practice, one considers that it can occur between $2m_Q$ 
and the threshold to produce open-heavy-flavour hadrons, $2m_{H}$. 
One subsequently multiplies this partial heavy-quark cross section by 
a phenomenological factor accounting for the probability that the
pair eventually hadronises into a given quarkonium state, $\P_{\cal Q}$. 
All this amounts to consider
\eqs{\sigma^{\rm (N)LO,\ direct/prompt}_{\cal Q}= \P^{\rm direct/prompt}_{\cal Q}\int_{2m_Q}^{2m_H} 
\frac{d\sigma_{Q\bar Q}^{\rm (N)LO}}{d m_{Q\bar Q}}d m_{Q\bar Q}.
\label{eq:sigma_CEM}}
$\P_{\cal Q}$ can be paralleled to the LDMEs in NRQCD, but for the fact that its
 size can be guessed if one assumes a simplified statistical-hadronisation scenario. 
Owing to the simplicity of the model, the direct or prompt yields are obtained 
from the same computation but with a different overall factor.

First, one expects~\cite{Amundson:1995em} that one ninth --one CS $Q\bar Q$ 
configuration out of 9 possible-- of the open-charm cross section in this 
invariant-mass region eventually hadronises into a ``stable''  quarkonium. In the
 case of $J/\psi$, beside this factor 9, it was argued~\cite{Amundson:1995em}, with
 a LO computation, that a simple statistical counting giving
\eqs{\label{eq:Fdir_CEM}
\P^{\rm direct}_{J/\psi} \overset{\hbox{\tiny stat. count.}}{\longrightarrow} 
\frac{1}{9} \frac{2 J_\psi +1}{\sum_i (2 J_i +1)} = \frac{1}{45},}
where the sum over $i$ runs  over all the charmonium states below the $D\bar D$ threshold, 
could describe the existing data in the late 90's.

NLO fits were then performed by Vogt in \cite{Bedjidian:2004gd} on data up to 
$\sqrt{s}=62$ GeV with $\P^{\rm direct}_{J/\psi}$  lying between 1.5~\% and 2.5~\%. 
This simple statistical counting rule works remarkably well for $J/\psi$, whereas 
it would not work for $P$-waves as discussed in~\cite{Lansberg:2006dh,Brambilla:2010cs} 
as well as for $\psi(2S)$. This shows the limit of the model in accounting for 
differences in the hadronisation of different quarkonium states. 
 
For the $\Upsilon$, the corresponding quantity is fit with a similar size, 
 between 2~\% and 5~\%. Following the state-counting argument, one would however expect 
a smaller number than for $J/\psi$. At this stage, it is important to reiterate that 
\ce{eq:Fdir_CEM} ignores phase-space constraints in the hadronisation process. What 
the CEM really predicts is  that $\P^{\rm direct}_{\cal Q}$ should be process independent.

\begin{figure}[hbt!]
\begin{center}
\includegraphics[width=11cm]{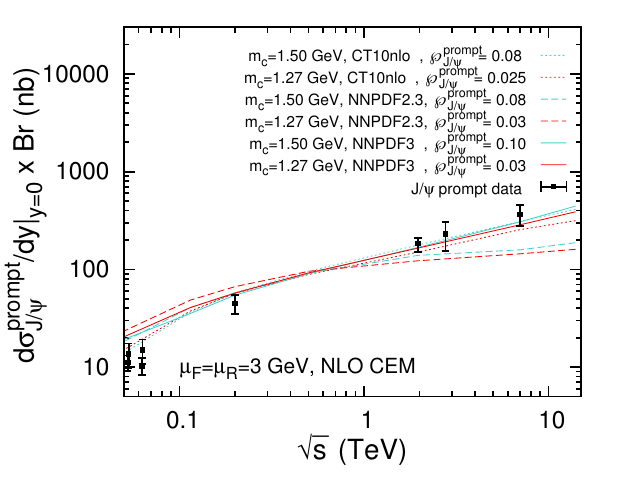}
\caption{Comparison between a selection of $P_T$-integrated prompt $J/\psi$ cross 
sections measured  at $y=0$ and the NLO CEM results  with two choices of the 
charm-quark mass ($m_c$) and of PDFs~\cite{Ball:2014uwa,Ball:2012cx,Lai:2010vv} as a function 
of $\sqrt{s}$ (see also~\cite{Feng:2015cba}). }\label{fig:sigma_CEM_vs_s}
\end{center}
\end{figure}

A NLO comparison with data from fixed-target as well as from colliders is shown 
on~\cf{fig:sigma_CEM_vs_s} for the $J/\psi$ case. 
Curves with different choices of the charm-quark mass and of the PDFs are shown. 
First we see that the mass dependence is nearly completely absorbed in the fit value of 
$\P^{\rm prompt}_{J/\psi}$. We stress that $m_Q$  impacts the evaluation of the
 cross-section principally via the range of integration over the pair invariant 
mass. Yet, the energy dependence is rather similar for both mass choices. One also sees 
the significant dependence on the PDF set. This is due to the rather low scale 
$\mu_R$ of this process and the very low $x$ values ($x \simeq M_\psi / \sqrt{s}$) 
reached. Such an effect does not appear at higher $x$ and at larger scales as for
$J/\psi +Z$. 

A thorough study of the connection between heavy-flavour and quarkonium 
production in the CEM can be found in~\cite{Nelson:2012bc}. Along the lines of 
this analysis, we will use $m_c=1.27$~GeV. Results with $m_c=1.5$ GeV are 
sometimes slightly different but never such as to modify the physics conclusion.

\subsection{The single-$J/\psi$ $P_T$ spectrum at NLO}

We turn now to the discussion of the $P_T$ spectrum of single $J/\psi$'s. As 
mentioned above, the CEM is expected to predict too hard a $P_T$ spectrum. Yet, 
in all Tevatron and LHC computations~\cite{Bedjidian:2004gd,Brambilla:2004wf}, 
the hard-scattering matrix element, which is employed, is at one loop for 
inclusive heavy-quark production, namely $\alpha_s^3$.  In fact, it is based on the 
well-known MNR computation~\cite{Mangano:1991jk} using the specific invariant-mass 
cut of \ce{eq:sigma_CEM}. At this order, the sole graphs contributing to the 
production of the heavy-quark pair (with or without invariant-mass cut) with a 
finite $P_T$ are those from $2\to 3$ processes. This de facto excludes any loop 
contribution. As such, all these existing computations at finite $P_T$ of the pair 
are effectively Born-order/tree-level computation from $gg [q \bar q] \to (Q \bar Q) g$
 or $gq \to (Q \bar Q) q$. It is therefore legitimate to wonder whether the 
resulting $P_T$ spectrum could be affected by large QCD $\alpha_s^4$ corrections, 
effectively NLO and not NNLO for this quantity. In particular, one could wonder 
whether the data can be better described at NLO and whether 
$\P^{\rm NLO}_{J/\psi}$ is different than  $\P^{\rm LO}_{J/\psi}$ ? 
In addition to be a significant advance in the CEM usage, such a computation is 
in fact relevant for our study since the $J/\psi$ measured by ATLAS with a $Z$ 
are in the range where the single-$J/\psi$ data starts to deviate from the CEM 
prediction at $\alpha_s^3$ and where such NLO corrections could matter both in 
the determination of the CEM parameter and in the hard-part computation for 
$J/\psi+Z$ itself. 

Given the straightforward connection between the CEM and heavy-quark production, 
such a computation is in fact not too demanding with modern tools of automated 
NLO frameworks, at the minimum cost of some slight tunings.  In particular, we 
have used {\small \sc MadGraph5\_aMC@NLO}~\cite{Alwall:2014hca} to perform our 
(N)LO CEM calculations for $J/\psi $ + a recoiling parton with a finite $P_T$ (and
 then of course for $J/\psi+Z$)\footnote{We stress that, for the CEM, there is no
 need to use specific automated tool like {\small \sc MadOnia}~\cite{Artoisenet:2007qm}
 and {\small\sc HELAC-Onia}~\cite{Shao:2012iz,Shao:2015vga} which are by the way
 currently restricted to tree level.}.  As already stated above, we have taken 
$m_c=1.27$ GeV, while we checked that $m_c=1.5$ GeV would only marginally change
 our results provided that the non-perturbative CEM parameter is chosen coherently.
 As what regards the parton distribution function (PDF), we have used the NLO 
NNPDF 2.3 PDF set~\cite{Ball:2012cx} with $\alpha_s(M_Z)=0.118$ provided by 
LHAPDF~\cite{Buckley:2014ana}. In this case, since the heavy-quark mass dependence 
is de facto absorbed in the CEM parameter, the main theoretical 
uncertainties\footnote{The uncertainty from the Monte Carlo integration may also 
be relevant for $J/\psi$ + parton production but only where the cross section is 
becoming very small.}  are coming from the renormalisation $\mu_R$ and 
factorisation $\mu_F$ scale variations which account for the unknown higher-order 
corrections. In practice,  we have varied them within 
$\frac{1}{2}\mu_0\le \mu_R,\mu_F \le 2\mu_0$ where the central scale $\mu_0$ 
is the transverse mass of the $J/\psi$ in $J/\psi$ + parton and the mass of $Z$ boson 
$M_Z$ in $J/\psi+Z$. It has been shown~\cite{Gong:2012ah} that the scale $M_Z$ 
is close to where the result is the most stable
for $J/\psi+Z$ at NLO in the CSM.

Since we wish to eventually use $\P^{\rm NLO,prompt}_{J/\psi}$ for our study of 
the ATLAS $J/\psi+Z$ data, we have restricted our study to the kinematical 
condition of the latest ATLAS single-$J/\psi$ data at $\sqrt{s}=8$ TeV~\cite{Aad:2015duc} 
corresponding to 11.4 fb$^{-1}$ and which are of course much more precise than 
previous data sets.

\begin{figure}[hbt!]
\begin{center}
\subfloat[LO]{\includegraphics[width=7.5cm]{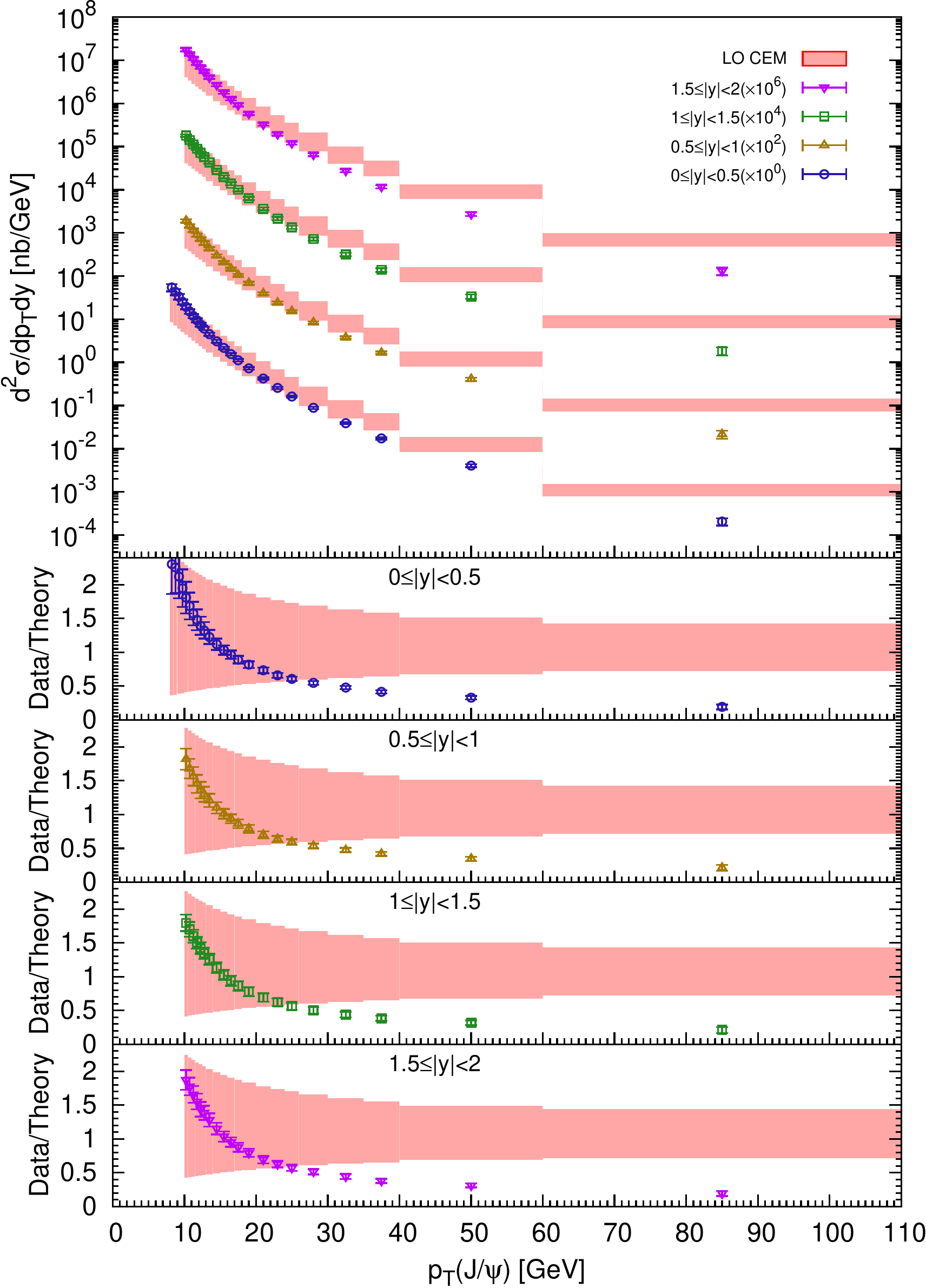}}
\subfloat[NLO]{\includegraphics[width=7.5cm]{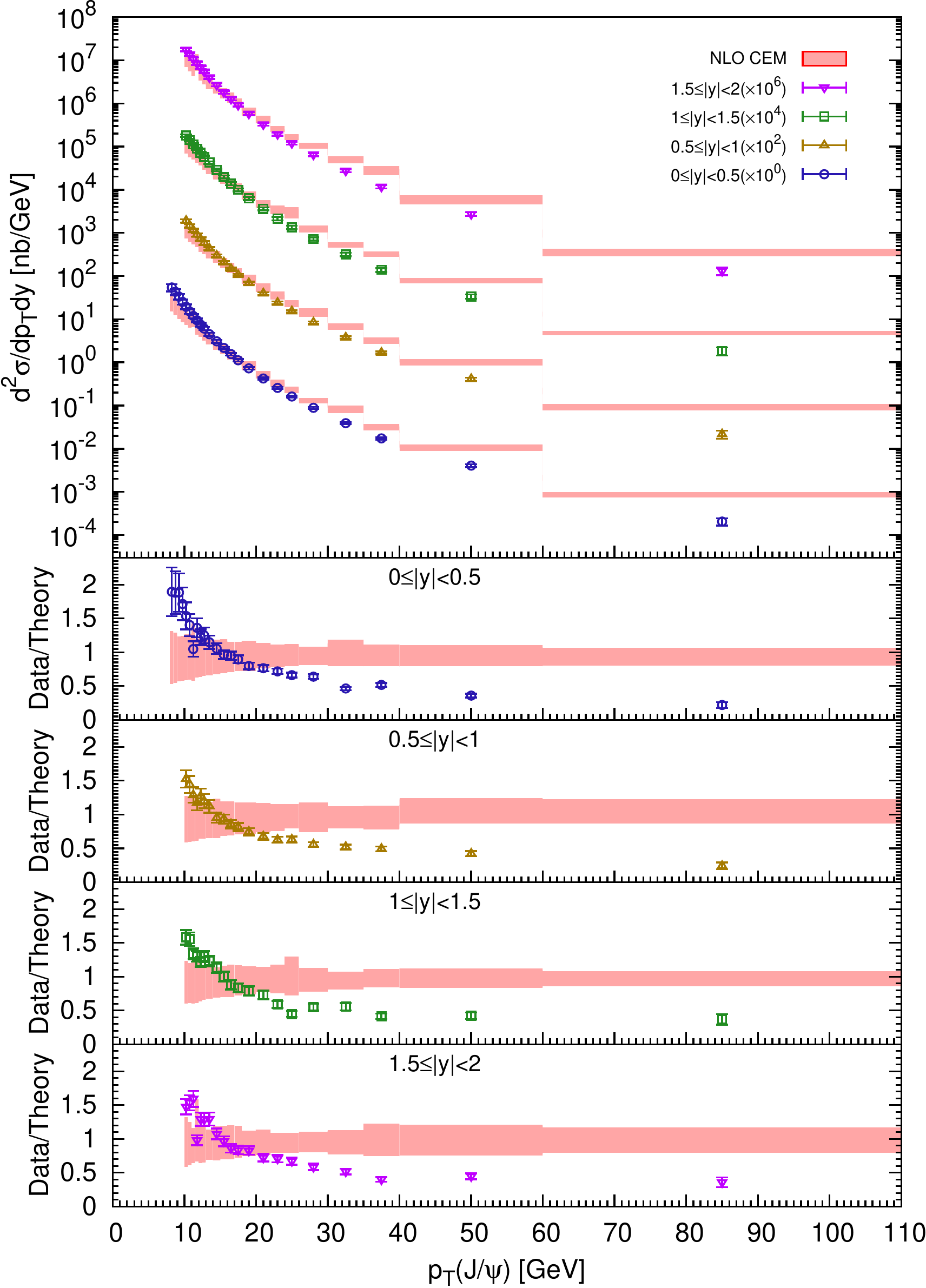}}
\caption{Comparison between 
the ATLAS data~\cite{Aad:2015duc} and the CEM results for $d\sigma/dy/dP_T$ of 
$J/\psi$ + a recoiling parton at (a) LO and (b) NLO at $\sqrt{s}=8$ TeV. [The 
theoretical uncertainty band is from the scale variation (see the text).] 
\label{fig:single-Jpsi-pT-spetrum}}
\end{center}
\end{figure}

Fitting this set with $m_c=1.27$ GeV, we obtain $\P^{\rm LO, prompt}_{J/\psi}=0.014\pm 0.001$ and
 $\P^{\rm NLO, prompt}_{J/\psi} =0.009 \pm 0.0004$. From these, we can deduce 
that the $K$ factor  affecting the $P_T$ slope is close to 1.6. As previously 
discussed, the CEM yields start to  depart from the data when $P_T$ increases 
(see \cf{fig:single-Jpsi-pT-spetrum}). This is happening in the region where 
fragmentation contributions are dominant as in $J/\psi+Z$ which makes us believe
 that the CEM will indeed provide an upper limit on SPS $J/\psi+Z$ computations. 

We note that the behaviour is similar at LO and NLO and that the uncertainties 
at NLO are smaller than at LO. We recall that the LO results (\ie\ for prompt 
$J/\psi$ + a recoiling parton) would coincide with the $P_T$ spectrum obtain 
from NLO code for inclusive prompt $J/\psi$ results~\cite{Bedjidian:2004gd,Brambilla:2004wf}.
This illustrates the added value of this first  NLO CEM study of the $P_T$
spectrum of single $J/\psi$ using a one-loop evaluation of $J/\psi$ + a recoiling 
parton. Similar studies for other quarkonia will be the object of a separate work.
For completion, let us add that the $\chi^2_{\rm dof}$ is 1.57 at NLO 
and 0.51 at LO, essentially because the LO scale uncertainties are 
larger. 

The value of $\P^{\rm (N)LO, prompt}_{J/\psi}$ is about a factor of 2--3 smaller 
than that obtained from the $P_T$-integrated total yields 
(see \cf{fig:sigma_CEM_vs_s}). The trend is opposed to that of NRQCD where the LDME 
fit values from the $P_T$-differential yield systematically overshoot that fit 
from the $P_T$-integrated total yields~\cite{Maltoni:2006yp,Feng:2015cba}. Thus even for the 
CEM, the universality $\P^{\rm prompt}_{J/\psi}$ seems to be challenged. 
In the following, we will naturally use the value of $\P$ fit to the 
$P_T$-differential yields.

\section{The $J/\psi+Z$ case\label{sec:result}}

\subsection{The ATLAS comparison with theory}

Let us now move on to the process of interest of this analysis, namely
$J/\psi+Z$ production at the LHC where the  $J/\psi$ is promptly produced, 
thus not from a $b$-hadron decay. 

Such a process has been studied in the past in NRQCD under 
the SPS mechanism. Only two NLO analyses exist, that of Ref.~\cite{Mao:2011kf} 
considered both CS and CO contributions, but strictly speaking is not a complete
 NLO NRQCD analysis since the $^3P_J^{[8]}$ transitions were disregarded. It 
would only be complete if the corresponding LDME was negligibly small but, as 
discussed above, this would contradict single-$J/\psi$ data. Following this 
study, another one considering only CS contributions appeared~\cite{Gong:2012ah}.
 It corrected a mistake in~\cite{Mao:2011kf}, presented first polarisation 
results and argued, owing to the scale dependence of the process, that a 
reasonable choice for the scales is rather $M_Z$ than $\sqrt{M_{J/\psi}^2+P_T^2}$ 
as used in~\cite{Mao:2011kf}.

When the ATLAS collaboration released its study, they attempted to compare
their data to these predictions. However they did not directly compare them to
their yield since they noted that 
a non-negligible part of the yield was probably from DPS contributions. 
This conclusion was motivated by two observations. First the distribution 
of the events as a function of the azimuthal angle between both detected particles,
$\Delta \phi$,
was showing a plateau close to 0, whereas a dominant SPS yield would show a peak 
at $\pi$, \ie\ for back-to-back events. Second, under the simple assumption that
the DPS yield comes from two uncorrelated scatterings, they evaluated it 
with the rudimentary pocket formula
\begin{eqnarray}
\sigma^{\rm DPS}(J/\psi+Z)=\frac{\sigma(J/\psi)\sigma(Z)}{\sigma_{\rm eff}}.
\end{eqnarray}
with $\sigma_{\rm eff}$ extracted from their $W+$ 2-jet analysis~\cite{Aad:2013bjm}, 
namely  $15 \pm 3 \hbox{(stat.)} ^{+5}_{−3} \hbox{(sys.)}$ mb and by applying their
cuts (see \ct{tab:phasespace}).
  Doing so, they assumed that a significant, but non-dominant, DPS yield was 
to be expected and that their yield could not simply be compared to the SPS
predictions above.  In order to extract the genuine SPS yield, 
the only one containing novel information on the quarkonium production 
mechanisms, they subtracted what they believed to be the DPS yield
 with $\sigma_{\rm eff} \simeq 15$ mb.

To be more precise, the data-theory comparison of~\cite{Aad:2014kba} was
done with the ratio of the $J/\psi+Z$ yield over that for $Z$ in order to reduce
the uncertainty from the $Z$ observation. The comparison with theory required 
them to evaluate $\sigma(Z)$ under their kinematical conditions 
(see \ct{tab:phasespace}), where they used $\sigma(pp\to Z \to e^+e^-)=533.4$ pb.
After the selection of the prompt $J/\psi$'s in the sample and the subtraction 
of the expected DPS yield (see above), they obtained :
\eqs{
^{\rm prompt}R^{\rm DPS\ sub}_{J/\psi+Z}&=\mathcal{B}(J/\psi\to\mu^+\mu^-)\,\frac{\sigma(pp\to Z+J/\psi)}{\sigma(pp\to Z)}\\
&=(45 \pm 13_{\rm stat} \pm 6_{\rm syst} \pm 10_{\rm DPS sub})\, \times\, 10^{-7}, 
}
whereas the CS based predictions are around $(1-5)  \times 10^{-7}$ and the most
 optimistic NRQCD-based predictions  (with CS and CO contributions) reaches
$9 \times 10^{-7}$. 

\begin{table*}[htpb]
\begin{center} \footnotesize
\begin{tabular}{c|c|c}
\hline\hline
\multicolumn{3}{c}{$Z$ boson selection}\\
\hline
\multicolumn{3}{c}{}\\
\multicolumn{3}{c}{$P_T$(trigger lepton)$>25$~{\rm GeV}, $P_T$(sub-leading 
lepton)$>15$~{\rm GeV}, $|\eta(\mathrm{lepton~from}~Z)|<2.5$}\\
\multicolumn{3}{c}{}\\
\hline\hline
\multicolumn{3}{c}{$J/\psi$ selection}\\
ATLAS fiducial~\cite{Aad:2014kba} & ATLAS inclusive~\cite{Aad:2014kba} & CMS 
fiducial~\cite{private-CMS}\\
\hline
$8.5<P_T^{J/\psi}<100\,{\rm GeV}$ & $8.5<P_T^{J/\psi}<100\,{\rm GeV}$ & $8.5<P_T^{J/\psi}<100\,{\rm GeV}$ \\
$|y_{J/\psi}|<2.1$ & $|y_{J/\psi}|<2.1$ & $|y_{J/\psi}|<2.1$\\
$P_T$(leading muon)$>4.0$~{\rm GeV} &  & $|\eta(\mathrm{muon})|<2.5$ \\
$|\eta(\mathrm{leading~muon})|<2.5$ & & \\
either \ldelim({2}{5mm}$P_T$(sub-leading muon)$>2.5$~GeV \rdelim){2}{1mm}[] & & \\
~~~~~~~~~~$1.3\leq |\eta(\textrm{sub-leading~muon})|<2.5$  & & \\
or ~~~~\ldelim({2}{5mm} $P_T$(sub-leading muon)$>3.5$~{\rm GeV} \rdelim){2}{1mm}[] &  &\\
~~~~~~~~~~$|\eta(\textrm{sub-leading~muon})|<1.3$ & & \\
\hline\hline
\end{tabular}
\caption{\label{tab:phasespace}Phase-space definition of the measured fiducial 
production cross-section following the geometrical acceptance of the ATLAS 
detector and the CMS detector.}
\end{center}
\end{table*}

As announced in our introduction,  the experimental results are indeed at least 
five times larger (corresponding to 2-$\sigma$ deviation) than the largest 
available theoretical predictions on the market. Larger LDMEs for the $^3S_1^{[8]}$ 
transition, in particular, could reduce the gap a little but at the cost of 
a discrepancy with the --much more precise-- single-$J/\psi$ data. In addition, 
introducing a nonzero $^3P_J^{[8]}$ contribution would probably interfere 
negatively with the dominant $^3S_1^{[8]}$ one. Instead of playing further with 
these parameters, we have thus decided to analyse the process in the CEM up to 
NLO accuracy, which provides an upper theory value above which any other similar
evaluations would probably be unrealistic, except from a totally overlooked 
partonic reaction. 

\subsection{NLO CEM SPS contributions and our DPS extraction}

The procedure to compute the $J/\psi+Z$ CEM cross section exactly follows from the 
same lines as for $J/\psi$ + a recoiling parton, where the hard scattering is 
$ij \to c \bar c + Z + k$ with $i$, $j$ and $k$  standing for $g$, $q$ or 
$\bar q$ (at LO, $k$ is irrelevant). The remaining of the procedure 
(invariant-mass cut, non-pertubative parameter, PDFs, scale variation, etc.) is 
exactly the same, although the central scale we take now is 
$\mu_0=M_Z$~\cite{Gong:2012ah} instead of the transverse mass of 
$J/\psi$~\cite{Mao:2011kf}.

To evaluate $^{\rm prompt}R^{\rm}_{J/\psi+Z}$, we used 
$\sigma(pp\to Z \to e^+e^-)=427$ pb\footnote{We have employed 
{\small \sc MadGraph5\_aMC@NLO}~\cite{Alwall:2014hca} to calculate 
$pp\rightarrow Z \rightarrow e^+e^-$ by taking into account the NLO QCD 
corrections. The spin-correlated decay $Z\rightarrow e^+e^-$ is done with 
the help of the module {\small\sc MadSpin}~\cite{Artoisenet:2012st}. We have 
also matched the NLO calculation of $pp\rightarrow Z \rightarrow e^+e^-$ to the 
parton showers provided by {\small\sc Pythia8.1}~\cite{Sjostrand:2007gs} via the 
MC@NLO method~\cite{Frixione:2002ik}. The cross section 
$\sigma(Z){\rm Br}(Z\rightarrow e^+e^-)$ in the $Z$ selection condition of 
Table.~\ref{tab:phasespace} is 427 pb at the fixed order NLO without spin 
correlations in the $Z$ decay. It varies by $20\%$ when the spin correlation and
 the parton-shower effects are taken into account, i.e. $505-520$ pb.}.
Our results for the NLO CEM SPS contributions are shown in \ct{tab:tot}. They 
are on the order of $8 \times 10^{-7}$ and similar to the most optimistic NRQCD ones 
but still  far for the ATLAS experimental value. It seems that the solution to 
the puzzle is not to be found in this direction. The $K$ factor for the hard 
part we found is 2.8 and the LO CEM yield (with $\mathcal{P}^{\rm LO,prompt}_{J/\psi}$) is
1.9 times smaller than the NLO yield (with $\mathcal{P}^{\rm NLO,prompt}_{J/\psi}$). 
The quark-gluon fusion channel at NLO is responsible for this large $K$ factor. 

\begin{table}[hbt!]
\begin{center}\small
\begin{tabular}{{c}*4 c} 
\hline\hline
 & exp  & LO CEM SPS & NLO CEM SPS  & DPS ($\sigma_{\rm eff}=4.70~{\rm mb}$) \\\hline
ATLAS inclusive~\cite{Aad:2014kba} & $63\pm13\pm5\pm10$ &$4.1^{+1.3}_{-1.0}$ &$7.6^{+2.0}_{-1.6}$ & $54.0$\\
ATLAS fiducial~\cite{Aad:2014kba} &  $36.8\pm6.7\pm2.5$ &$2.2^{+0.7}_{-0.6}$ &$4.2^{+1.1}_{-0.9}$ & $22.6$\\
CMS fiducial & --  & $3.9^{+1.3}_{-0.9}$    &$7.5^{+2.0}_{-1.6}$ & $52.5$\\
\hline\hline
\end{tabular}
\caption{\label{tab:tot}Comparison for  
the cross-section ratio $^{\rm prompt}R_{J/\psi+Z}$ between the CEM predictions 
for the SPS yields, our DPS extraction and the experimental results at 
$\sqrt{s}=8$ TeV. The theoretical uncertainty for the NLO SPS is from the 
renormalisation and factorisation scales. [All quantities are in units of $10^{-7}$].}
\end{center}
\end{table}

\begin{figure}[hbt!]
\begin{center}
\includegraphics[width=10cm]{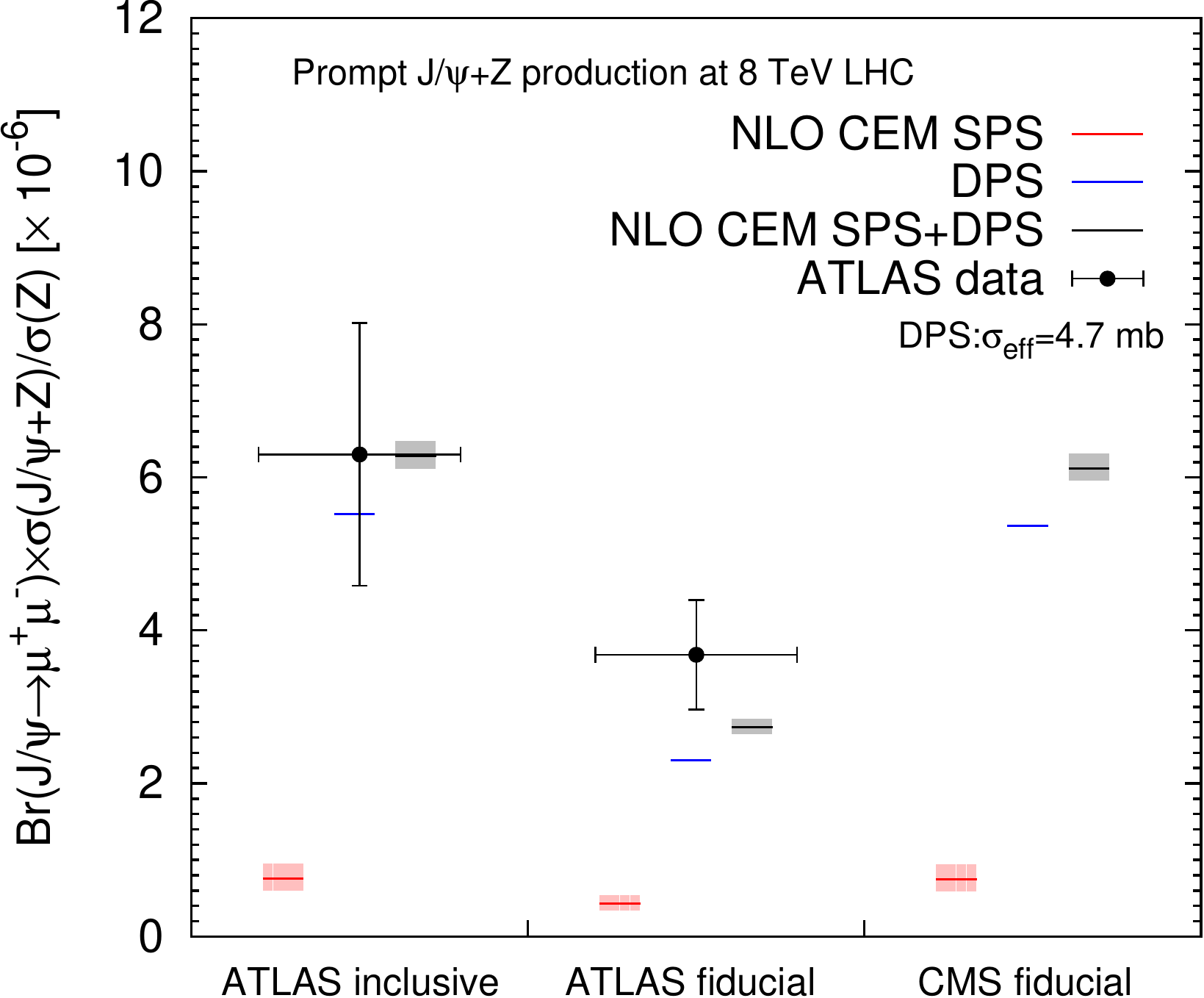}
\caption{Summary of the total yields of prompt $J/\psi+ Z$ production at 
$\sqrt{s}=8$ TeV at the LHC. ATLAS data~\cite{Aad:2014kba} (both "inclusive" and
 "fiducial") are shown for comparison.}\label{fig:sigma_tot}
\end{center}
\end{figure}

Apart from questionning the reliability of the ATLAS measurements or from 
invoking new physics contributions, the only other possibility left to solve 
the gap is to question the size of the DPS yield subtracted by ATLAS. As just 
said, they opted for a $\sigma_{\rm eff}$ close to 15 mb based on their 
$W+$ 2-jet analysis. Yet, recent quarkonium-related 
analysis~\cite{Lansberg:2014swa,Abazov:2014qba,ATLAS-CONF-2016-047} have pointed at 
values smaller than 10 mb, which would in turn induce a larger DPS yield to be subtracted. 

We have therefore decided to simply fit $\sigma_{\rm eff}$ to the total "inclusive" 
ATLAS $J/\psi+Z$ yield along with the NLO CEM one for the SPS contribution and 
we obtained $\sigma_{\rm eff} = 4.7$ mb (see \ct{tab:tot} 
and \cf{fig:sigma_tot}).  Although, in principle, we could have 
followed the procedure we used for $J/\psi+J/\psi$ production 
in~\cite{Lansberg:2014swa} to estimate the DPS contribution, we have decided to rely 
on the ATLAS computation, apart from the normalisation obviously. In other 
words, our DPS yield is 3 times larger than the one employed by ATLAS, \ie\ 
corresponding to a $R$ of $54 \times 10^{-7}$  vs $18 \times 10^{-7}$. In order 
to have an estimation of the DPS yield for the ATLAS and CMS fiducial regions, 
which are not given in~\cite{Aad:2014kba}, we have used a Crystal Ball function 
fit (see~\cite{Lansberg:2014swa} for technicalities) to calculate the acceptance 
of $J/\psi$ production from DPS. This allows us to compare our results to the 
ATLAS measurement in their fiducial region and to predict the total yields 
within the CMS fiducial region. 

Using these numbers, we can also derive an upper limit on $\sigma_{\rm eff}$ (corresponding
to the smallest acceptable DPS yield) by subtracting the 1-$\sigma$ higher value
of the NLO CEM yield from the 1-$\sigma$ lower value of the ATLAS measurements. 
This gives 7.1 mb. Even combining both these extreme limits, the ATLAS yield points
at a somewhat large DPS. If now one assumes no SPS at all, one can extract a lower value
for $\sigma_{\rm eff}$ as low as 3.2 mb. 
 
\subsection{The $P^{J/\psi}_T$ dependence}

Obviously, a DPS yield three times larger (with $\sigma_{\rm eff} =4.7$ mb)  
would significantly change the azimuthal distribution from which the ATLAS 
collaboration concluded for a significant, but non-dominant DPS yield. Under our
 assumption, it becomes now the dominant contribution, about five times as large 
as the SPS one. Before showing that it does not create any tension with this 
azimuthal distribution,  we however need to discuss the cross section of a 
function of $P^{J/\psi}_T$, for a reason which will become clear in the next section.

\begin{figure}[hbt!]
\begin{center}
\subfloat[ATLAS inclusive]{\includegraphics[width=7cm]{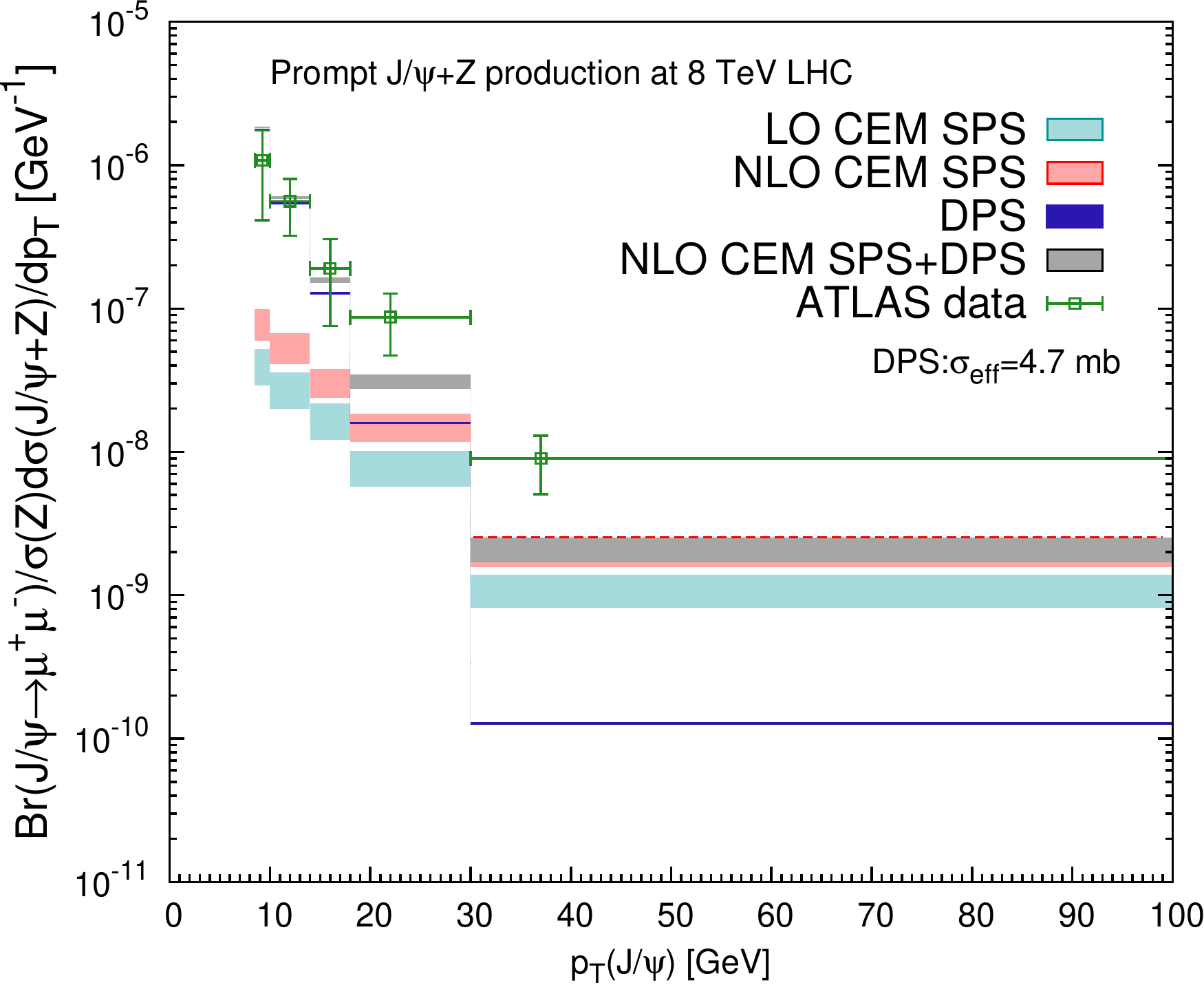}}
\subfloat[CMS fiducial]{\includegraphics[width=7cm]{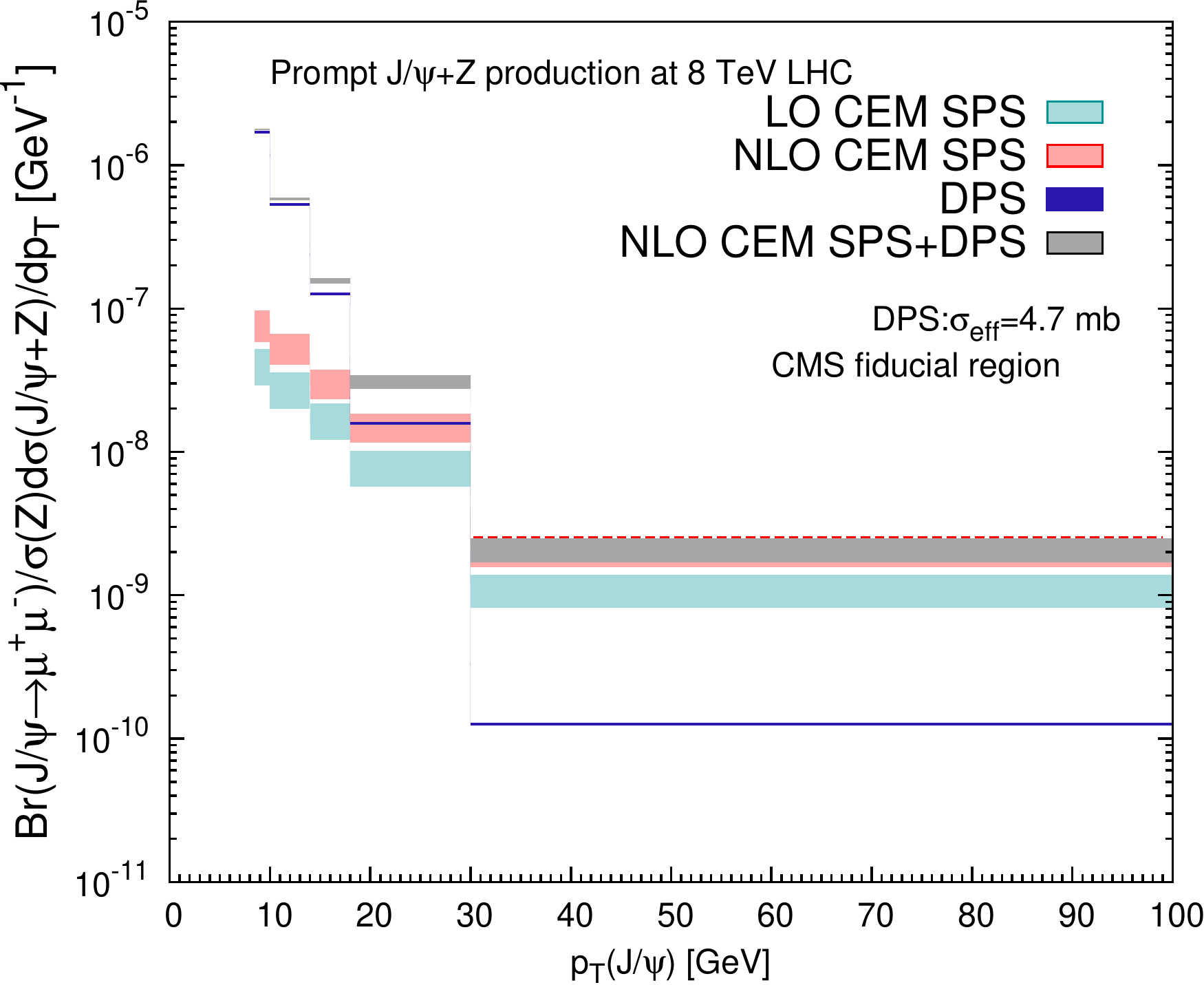}}
\caption{The $J/\psi$ $P_T$ dependence of $R$: (a) comparison between the ATLAS 
data~\cite{Aad:2014kba}, the CEM results for $J/\psi+Z$  at NLO (and LO) and our 
extracted DPS yield in the ATLAS acceptance at $\sqrt{s}=8$ TeV; (b) predictions
 in the CMS acceptance.
\label{fig:dRdpsipt}}
\end{center}
\end{figure}

As what regards the SPS CEM cross section, the $P^{J/\psi}_T$ spectrum 
is computed exactly as above with just  one less integral, whereas 
the $P_T$ dependence of the DPS yield is exactly the one of ATLAS from 
Table 5 of~\cite{Aad:2014kba} with a simple rescaling
of the normalisation due to the change in $\sigma_{\rm eff}$.
The resulting dependence (in form of the ratio $R$) is shown on \cf{fig:dRdpsipt}.
The sum of the SPS and DPS yield in gray gives a reasonable account of the ATLAS yield, 
with a slight gap opening at large $P_T$. The fact that the agreement is good at 
low $P_T$ is however just a consequence of the fit of $\sigma_{\rm eff}$. 
Such a distribution is rather a consistency check than a test. 

However, this helps to clearly illustrate how the low $P^{J/\psi}_T$ yield is 
heavily dominated by the DPS contributions (as is the total yield) and that 
the high $P^{J/\psi}_T$ yield is exclusively from SPS contributions. This a 
key observation for our discussion in the next section.

\subsection{The azimuthal distribution}

This difference of the $P^{J/\psi}_T$ spectrum indeed has an unexpected 
consequence on the interpretation of the azimuthal distribution of the ATLAS 
events since it was done with event counts, without efficiency 
correction\footnote{The efficieny was however checked to be constant in 
$\Delta \phi$~\cite{private-Darren}.}. It happens that the ATLAS efficiency is 
much higher for the last bin in $P^{J/\psi}_T$ than for 
the first bin, up to 3 times in fact. 

This is visible from the statistical uncertainties in \cf{fig:dRdpsipt} 
which remain
more or less constant whereas the yield is admittedly much smaller in the last bin.
In turn, the events used for the  $\Delta \phi$ distribution results from a  biased
sample which is strongly enriched in high $P^{J/\psi}_T$ events. As we discussed in 
the previous section, high $P^{J/\psi}_T$ events are essentially of SPS origin thus 
mostly populating the  $\Delta \phi \sim \pi$ region. Our claim is that the fact 
that the peak is visible is only due to that and not to a large SPS yield in general. 
In other words, such a distribution (unless corrected for efficiency in the future) 
{\it cannot} be used to discuss the ratio DPS/SPS integrated in 
$P^{J/\psi}_T$, nor to reliably extract the DPS yield or $\sigma_{\rm eff}$. 
The lower limit obtained by ATLAS DPS by assuming that the first bin in 
$\Delta \phi \sim \pi$ was fully from DPS events, 5.3 mb, is  slightly different 
than ours since it ignores the information from data at $\Delta \phi$ 
away from 0. Without proper information on the DPS/SPS ratio in each
bin (which can be process dependent\footnote{For process like 
$J/\psi$-pair production, it is well know that it very much depends 
on the $P_T$ cuts \cite{Kom:2011bd,Lansberg:2013qka,Baranov:2015cle}.}), this 
remains an assumption to consider that the $\Delta \phi \sim  0$ bin is entirely 
fed by DPS events.

\begin{table}
\begin{center}
\begin{tabular}{{c}*2 c} 
\hline\hline
 $P_T^{J/\psi}$ [GeV] & $S$\\\hline
$(8.5,10)$ & $10.6$\\
$(10,14)$ & $21.0$\\
$(14,18)$ & $6.2$\\
$(18,30)$ & $9.8$\\
$(30,100)$ & $8.6$\\
$(8.5,100)$ & $56.1$\\
\hline\hline
\end{tabular}
\caption{\label{tab:events}The estimation of the number of the signal events $S$
 (before the efficiency corrections) in each $P_T^{J/\psi}$ bin with the 
assumption $B/S=17/P_T^{J/\psi}$.}
\end{center}
\end{table}

\begin{figure}[hbt!]
\begin{center}
\subfloat[LO]{\includegraphics[width=7cm]{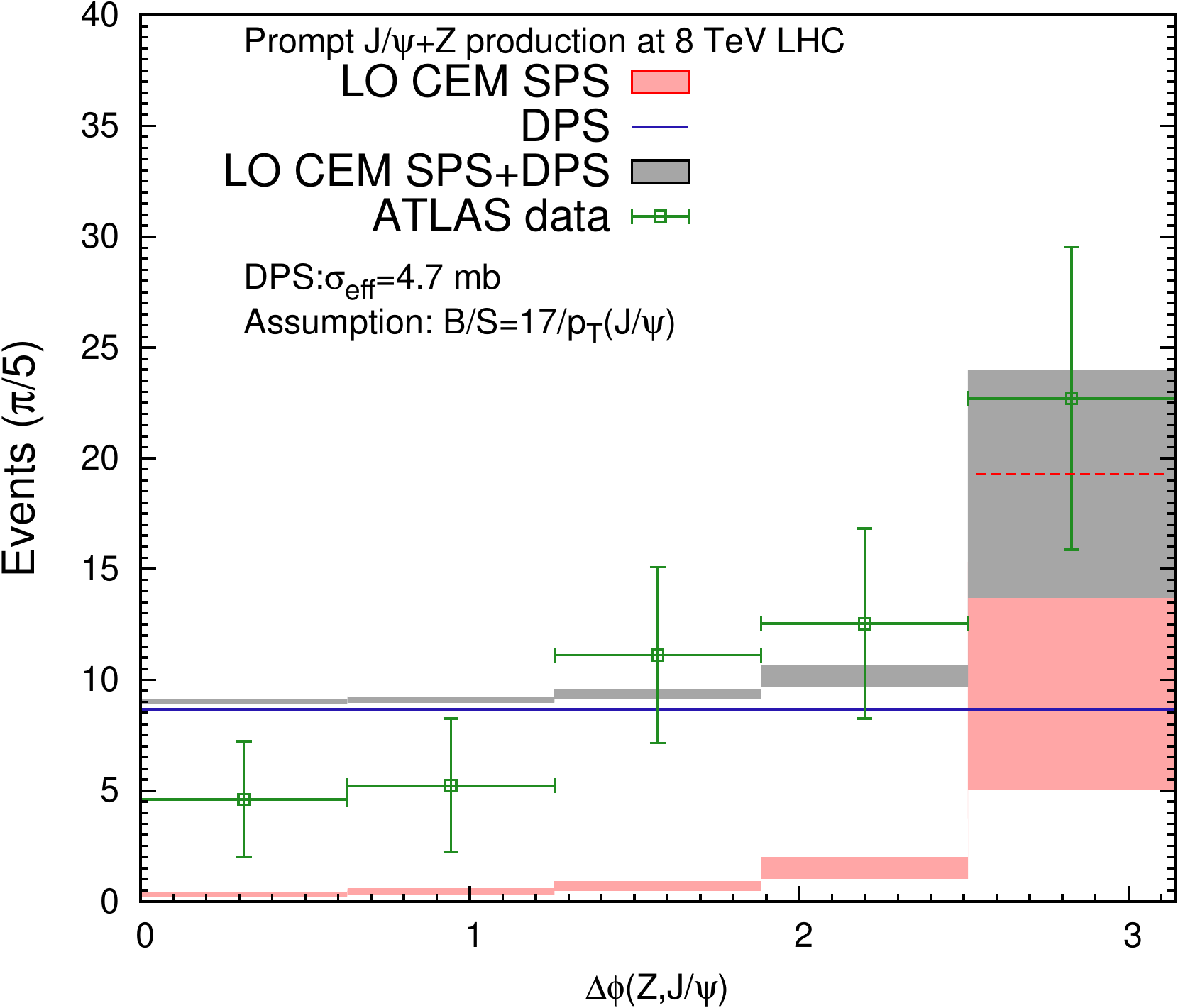}}
\subfloat[NLO]{\includegraphics[width=7cm]{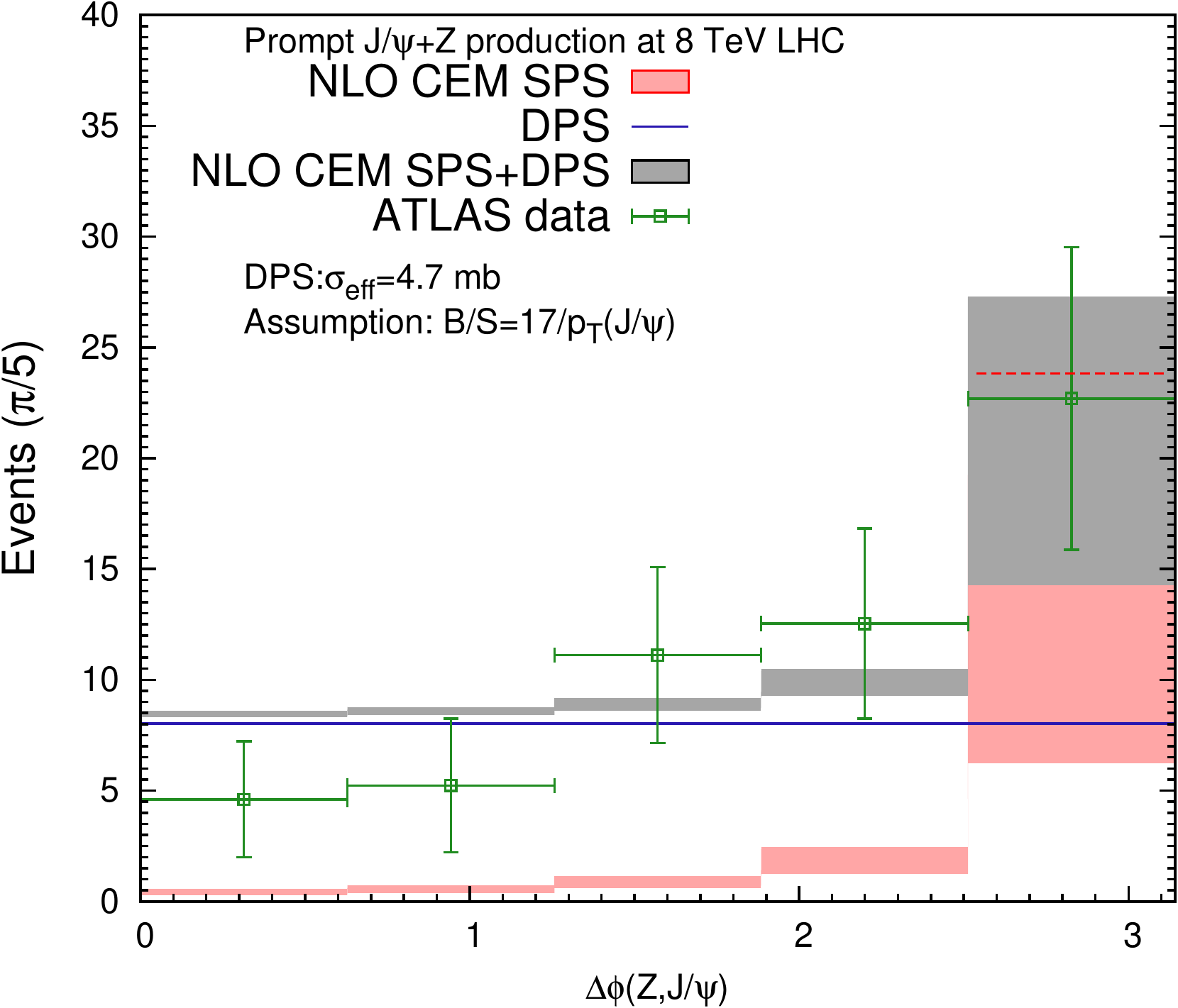}}
\caption{Comparison between the (uncorrected) ATLAS azimuthal event distribution
 and our theoretical results for $J/\psi+Z$  at (a) LO (resp. (b) NLO) CEM SPS + DPS
effectively folded with an assumed ATLAS efficiency.
\label{fig:dphi}}
\end{center}
\end{figure}

A way to check our hypothesis is to approximately fold our DPS and SPS $P^{J/\psi}_T$ spectra with 
the ATLAS efficiency, to plot and sum the DPS and SPS $\Delta \phi$ distributions.
Since the ATLAS efficiency as a function of $P^{J/\psi}_T$ is not publically released, 
we have used the following simple makeshift, \ie\ to derive its $P^{J/\psi}_T$
dependence from the corrected yield dependence (from  \cf{fig:dRdpsipt}) 
and the raw number of events in each bins derived from the statistical uncertainty 
quoted by ATLAS (see also~\cf{fig:dRdpsipt}).  Such an estimation however also 
requires the knowledge of the  background  size, $B$, which was not given, but 
which we assumed to be suppressed with respect to $S$ (\ie\ the true $J/\psi+Z$ 
events) as $B/S \propto (P^{J/\psi}_T)^{-1}$.

Approximately knowing $S$ (see \ct{tab:events}) and the theoretical SPS/DPS fraction
 in each $P^{J/\psi}_T$ bin, we were able to  stack our theoretical events SPS 
and DPS events, bin by bin in $P^{J/\psi}_T$, in the $\Delta \phi$ plot with 
their specific $\Delta \phi$ distributions (flat for DPS, peaked at 
$\Delta \phi \simeq \pi$ for the SPS following our NLO CEM computation).
The resulting distribution is shown on \cf{fig:dphi}. They look essentially 
the same at LO and NLO and demonstrates that increasing the DPS yield by a 
factor of 3 does not create any tension with the observed ATLAS {\it event} 
$\Delta \phi$  distribution if the efficiency corrections are approximately 
accounted for in the theory evaluations. We are of course eager for an updated 
analysis by ATLAS to avoid such a complicated procedure to compare the theory to 
the experimental $\Delta \phi$  distribution.

\subsection{The $P^{Z}_T$ dependence}

Before concluding, we also give our predictions for the $P^{Z}_T$ dependence of 
the yield. For the SPS yield, it follows from our NLO CEM evaluation. For the DPS 
yield, it follows from the simple observation that the absence of correlation 
between the $J/\psi$ and the $Z$ kinematics in DPS events\footnote{Such a 
statement remains sound as long as the initial parton-momentum fractions $x$ are 
far from 1/2.} allows us to state that the normalized distribution 
$\frac{1}{\sigma}\frac{d\sigma}{dP^{Z}_T}$ should be  identical for both the 
single-$Z$ production and the (DPS) $J/\psi+Z$ production. In fact, this could be 
used to check the dominance of the DPS yield in some part of the phase space. 
The normalised distribution of $\frac{1}{\sigma}\frac{d\sigma}{dP^{Z}_T}$ for 
inclusive $Z$ production can accurately be predicted with the help of 
{\small \sc MadGraph5\_aMC@NLO}, which can be evaluated at NLO and matched to 
parton-shower routines.

\cf{fig:dRdZpt} shows the $P^{Z}_T$ dependence of $R$ for both the ATLAS and CMS 
acceptances. One notices that the NLO and LO SPS spectra are different for large 
$P^{Z}_T$ values. This shows that the leading topologies to generate high-$P_T$ 
$Z$ bosons are not present at LO. One also observes that the DPS yield dependence 
is softer than the NLO SPS one and similar to the LO SPS one.

\begin{figure}[hbt!]
\begin{center}
\subfloat[ATLAS inclusive]{\includegraphics[width=7cm]{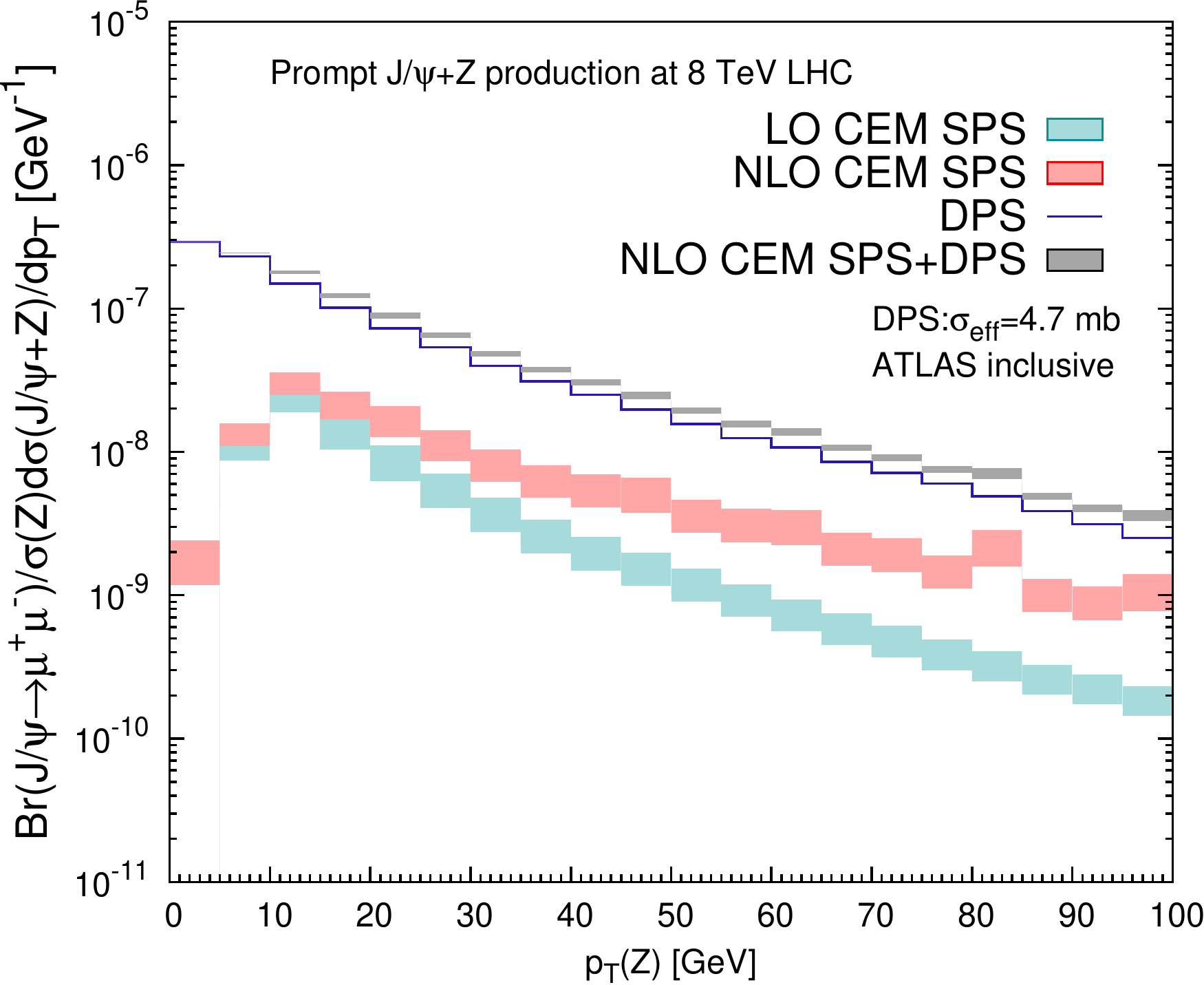}}
\subfloat[CMS fiducial]{\includegraphics[width=7cm]{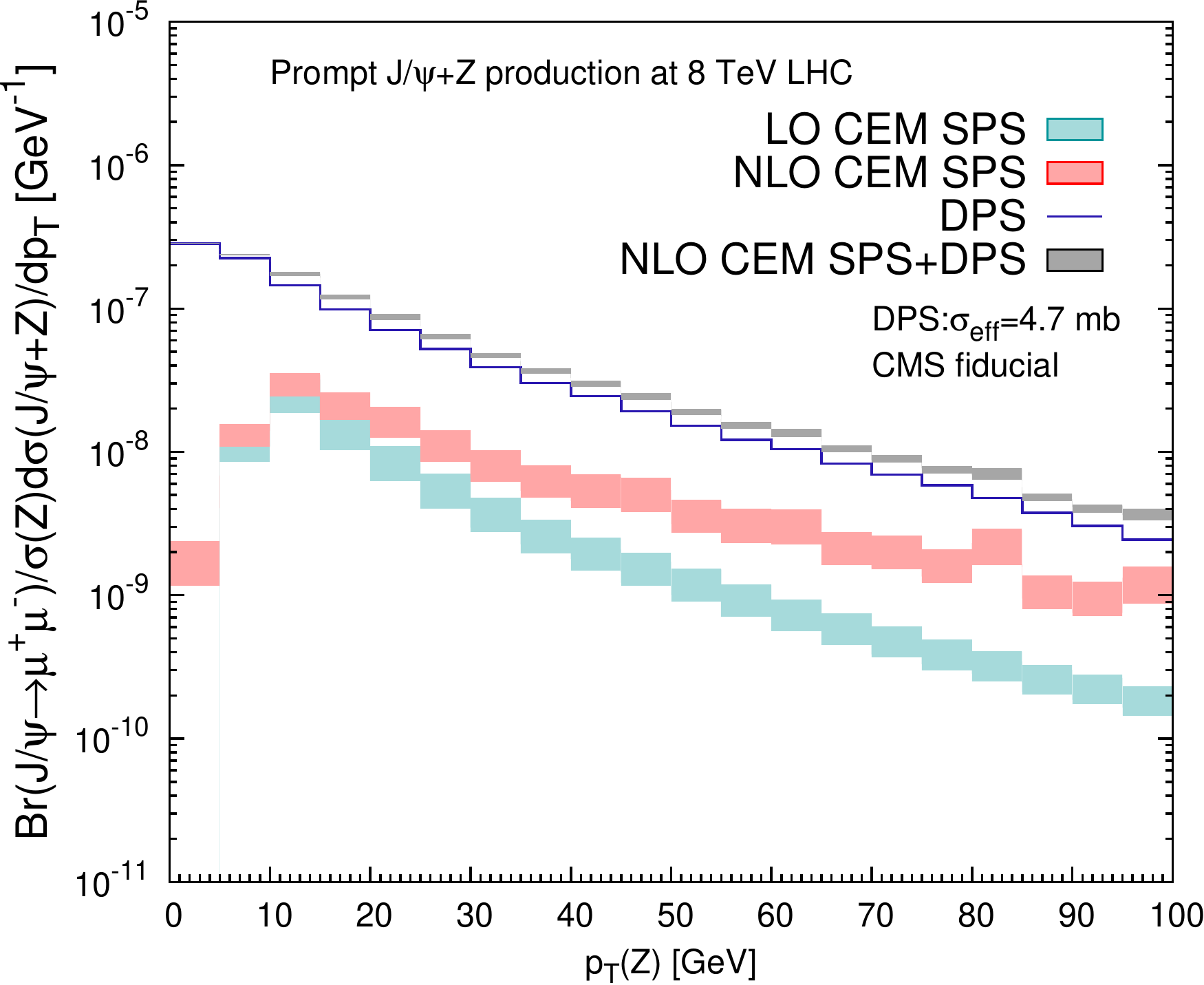}}
\caption{Our (N)LO CEM SPS+DPS predictions of prompt $J/\psi+Z$ production at 
$\sqrt{s}=8$ TeV LHC in the (a) ATLAS inclusive and (b) CMS fiducial regions.
\label{fig:dRdZpt}}
\end{center}
\end{figure}

\section{Conclusions\label{sec:conclusion}}

In this paper, we have performed a theoretical re-analysis of the associated 
production of a prompt $J/\psi$ with a $Z$ boson at the LHC in view of the 
thought-provoking results of ATLAS~\cite{Aad:2014kba} at $8$ TeV LHC. Despite 
its cross section on the order of a few femtobarns, this process has the 
potential to tell us much about the quarkonium-production mechanisms and on the 
DPS physics in the case of a highly asymmetric system.  

For the first time, we have performed a NLO calculation of the hadroproduction 
of a $J/\psi$ + a recoiling parton in the CEM with the help of {\small \sc 
MadGraph5\_aMC@NLO} and fixed the CEM non-perturbative parameter $\P^{\rm NLO, 
prompt}_{J/\psi}$ by fitting the most recent and precise ATLAS data. With the
extracted $\P^{\rm NLO, prompt}_{J/\psi}$, we have presented the first 
theoretical calculation of the prompt $J/\psi+Z$ SPS production at NLO in the 
CEM which directly follows from the quark-hadron-duality principle in the 
context of quarkonium production. We do believe that our calculation can be 
considered as a conservative upper value of the SPS yield. In other words, any 
NRQCD evaluation with a larger yield would likely be based on parameters 
incompatible with the existing single-$J/\psi$ data. This allows us to state that, 
in the ATLAS and CMS fiducial regions\footnote{Note that the validity of 
such a statement can depend on the acceptance cuts, especially on the fact that, 
in the ATLAS and CMS acceptances, low $P_T$ $J/\psi$'s
are rejected but not the low $P_T$ $Z$'s.}, SPS contribution dominate the "high" 
$P_T^{J/\psi}$ region, where the DPS contributions tend to dominate the "low" 
$P_T^{J/\psi}$ one. Complementary measurements by LHCb at $P_T^{J/\psi}$ as low 
as a few GeV would therefore be extremely important whereas future measurements 
at higher $P_T$, with larger luminosities, would normally test the 
quarkonium-production mechanisms.

A thorough comparison with the ATLAS data has then been made by taking into account both 
SPS and DPS contributions. We argue that the DPS yield subtracted by ATLAS with 
 $\sigma_{\rm eff}\simeq 15$~mb was underestimated and thus the expected impact 
of SPS overestimated. In fact, their fitted
lower value, 5.2 mb, assuming the absence of SPS contributions at 
$\Delta \phi \to 0$ and ignoring their $\Delta \phi$ event distribution, is 
well compatible with our argument. With our NLO SPS results, we are indeed able to 
fit $\sigma_{\rm eff} \simeq 4.7$ mb from the $P_T$-integrated total yield and 
compute the other differential distributions (e.g. azimuthal distribution and 
transverse momentum distributions). Considering the CEM as indeed an upper 
theory limit for the SPS yield, we are also able to derive an upper value for 
$\sigma_{\rm eff}$ corresponding to the smallest possible DPS yield in agreement
 with the data. On the other hand, assuming a negligible SPS contribution to the 
$P_T$-integrated yield\footnote{hypothesis however disfavoured by differential 
distributions.}, we obtain a virtual lower $\sigma_{\rm eff}$ limit, which is 
 more conservative than the one derived by ATLAS. The $\sigma_{\rm eff}$
 range obtained likewise is in good agreement with other quarkonium-related 
extractions (see \cf{fig:sigma_eff}) and is visibly lower than the ones extracted from 
jet-related observables, pointing at a possible process dependence of 
$\sigma_{\rm eff}$. Indeed, one should keep in mind that all these extractions 
rely on the implicit factorisation of the pocket formula ($\sigma_{A+B}^{\rm DPS} \propto
\sigma_A \times \sigma_B$) and there does not exist proofs of such a 
factorisation. In fact, factorisation-breaking effects have been discussed in a 
number of recent studies (see \eg~\cite{Blok:2013bpa,Kasemets:2012pr,Diehl:2014vaa}).

\begin{figure}[hbt!]
\begin{center}
\includegraphics[width=10cm]{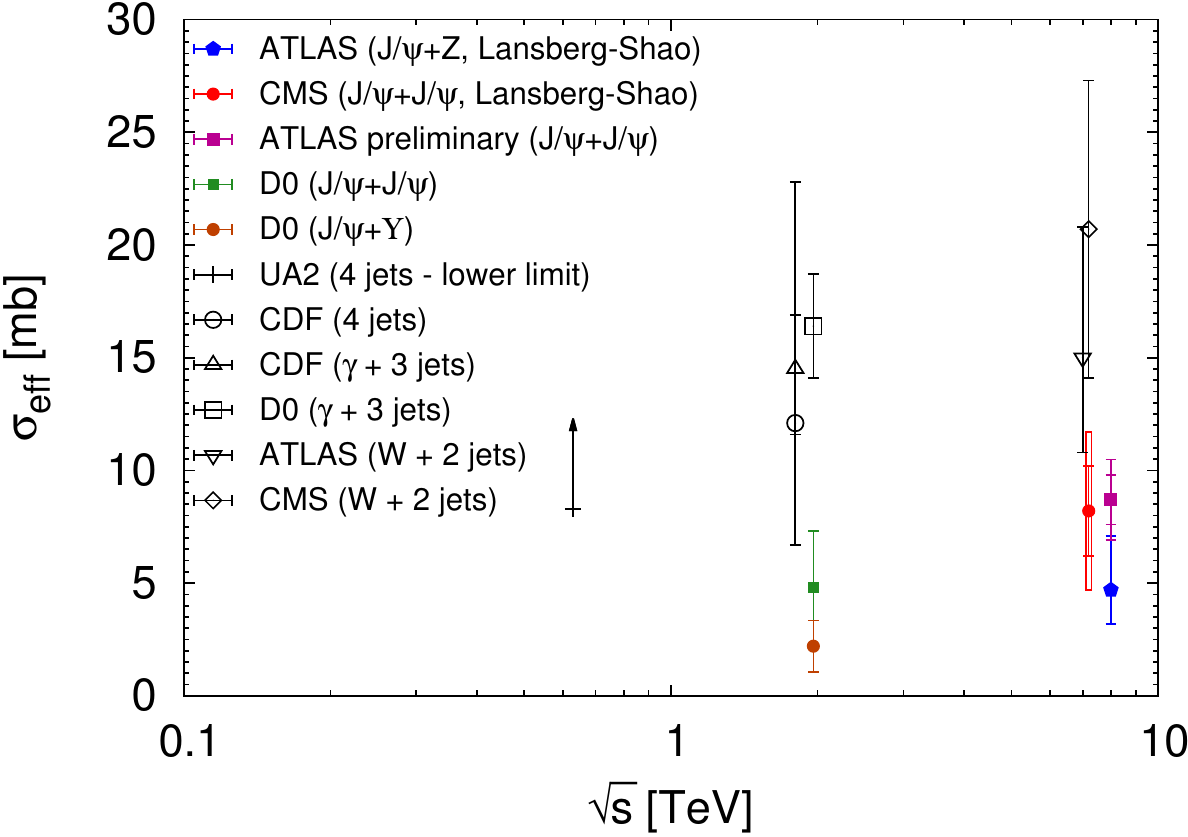}
\caption{Comparison of our range for $\sigma_{\rm eff}$ ($4.7^{+2.4}_{-1.5}$~mb)
 extracted from the $J/\psi+Z$ data with 
other extractions~\protect\cite{Akesson:1986iv,Alitti:1991rd,Abe:1993rv,Abe:1997xk,Abazov:2009gc,Aad:2013bjm,Chatrchyan:2013xxa,Abazov:2014qba,Lansberg:2014swa,ATLAS-CONF-2016-047}.\label{fig:sigma_eff}}
\end{center}
\end{figure}

These results are quite encouraging as the agreement between our theoretical 
results and the ATLAS data is acceptable, given the reduced number of events, and only a small discrepancy is present 
in $P_T^{J/\psi}$ distribution. This calls for further analyses by experiments. 
In this paper, we also provide our predictions for the ongoing CMS measurement. 
We are also hopeful that our theoretical evaluation will motivate further theory
 updates at NLO to substantiate our claim that the DPS-subtracted yield of ATLAS
 can indeed be accounted by known mechanisms.

\section*{Acknowledgements} 
We thank Z. Conesa del Valle, V. Kartvelishvili, Y. Kubota, S. Leontsinis, 
F. Maltoni, D. Price, L.-P. Sun, J.~Turkewitz for useful discussions. The work 
of J.P.L. is supported in part by the French CNRS via the LIA FCPPL 
(Quarkonium4AFTER) and  the D\'efi Inphyniti-Th\'eorie LHC France. H.S.S. is 
supported by the ERC grant 291377 \textit{LHCtheory: Theoretical predictions 
and analyses of LHC physics: advancing the precision frontier}.

\bibliographystyle{utphys}

\bibliography{onium+Z-CEM-090816}

\appendix

\end{document}